\documentclass[prl,aps,twocolumn,showpacs,preprintnumbers,amsmath,amssymb]{revtex4}
\usepackage{epsf}

\begin{document}
\draft
\newcommand{\ve}[1]{\boldsymbol{#1}}

\title{Interface controlled electronic charge inhomogeneities in correlated heterostructures}
\author{Natalia Pavlenko\cite{address2} and Thilo Kopp}
\address{Center for Electronic Correlations and Magnetism, Universit\"at Augsburg, Germany}

\begin{abstract}
For heterostructures of ultrathin, strongly correlated copper-oxide films and
dielectric perovskite layers, we predict inhomogeneous electronic interface
states. Our study is based on an extended Hubbard model for the cuprate film.
The interface is implemented by a coupling to the electron and phonon degrees of freedom
of the dielectric oxide layer. We find that electronic ordering in the
film is associated with a strongly inhomogeneous
polaron effect. We propose to consider the interfacial tuning
as a powerful mechanism to control the charge ordering in
correlated electronic systems.
\end{abstract}

\pacs{74.81.-g,74.78.-w,73.20.-r,73.20.Mf}


\maketitle

The surfaces and interfaces in strongly correlated transition metal oxides are
of fundamental importance for artificial complex oxide
superlattices and oxide electronic devices. Moreover they open a new field
to study correlation effects not observed in the bulk.
In the Mott insulating states of correlated oxides, the
interplay of charge, spin and lattice degrees of freedom combined with the
chemical doping by electrons or holes, leads to a wealth of complex behavior,
as in the high-T$_c$ cuprates and CMR-manganites
\cite{tokura_dagotto}. Here, the doped carriers, giving rise to superconductivity
and colossal magnetoresistance, also lead to charge ordered states
\cite{abbamonte}. In manganites like La$_{2-2x}$Sr$_{1+2x}$Mn$_2$O$_7$ or
La$_{0.5}$Sr$_2$MnO$_4$, order at certain doping
levels is characterized by superstructures of Mn$^{3+}$ and Mn$^{4+}$ ions in
MnO$_2$-planes \cite{tokura_dagotto}.
Clusters with local order appear near the
ferromagnetic Curie temperature where the CMR-effect is observed. On the other
hand, in cuprates, recent STM studies indicate hidden electronic
ordered states like checkerboard four CuO$_2$ unit cell patterns or stripes
\cite{hoffman}. The checkerboard states
\cite{davis,hoffman,vershinin} coexist with the superconducting state and
possibly play an important role in the appearance of superconductivity.

As compared to surface-related states, the physics is even more complex and less understood
at interfaces between correlated oxides and structurally compatible perovskites.
If the heterostructure is formed by an oxide film grown on an insulating layer with high
dielectric constant like SrTiO$_3$ or BaTiO$_3$, the correlated charge carriers of the
film are coupled at the interface to the charge and lattice degrees of freedom
of the dielectric layer. Recent studies of complex oxide interfaces show
that not only chemical or electrostatic doping, but also the microscopic
structure of the interface perovskite titanate layer can lead to a
metal-insulator transition with a mixed valence charge modulation in
LaTiO$_3$/SrTiO$_3$ superlattices \cite{ohtomo,millis} or to a suppression of
the superconducting state in
YBCO/SrTiO$_3$ heterostructures \cite{eckstein,pavlenko_kopp}. In the
perovskite titanate layers, the nonzero dielectric polarization is stabilized
due to the hybridization between Ti$^{4+}$ $d$- and O$^{2-}$
$p$-orbitals in the TiO$_6$--octahedra \cite{cohen}. The TiO covalent
hybridization is strongly modulated by the dynamical polar lattice
displacements \cite{migoni} which is essential for ferroelectricity in
perovskite titanates. Consequently, if the strongly correlated oxide film of a few nm thickness
is grown on a dielectric layer, the new interface states caused by the TiO
dynamical covalent charge are not shunted by ``bulk behavior'' and dramatically
affect the properties of the entire system.

In this work, we consider a basic model which allows to investigate how the
dynamical covalency and the phonons of the perovskite titanate can act on the
correlated states of the oxide film.
For the oxide films, formed by
weakly-coupled transition metal oxide planes, we focus on the first plane at
the interface without detailed analysis of the interplanar charge
redistribution.
With regard to recent experiments \cite{davis,hoffman}, we
allow for electronic states with a checkerboard ordering.
We consider an extended Hubbard-type model on a $N$-site square lattice
with tight-binding dispersion $\varepsilon_{\ve{k}}=-2t
\eta_{\ve{k}}^+$, $\eta_{\ve{k}}^+=\cos k_x+\cos k_y$:
\begin{eqnarray}\label{hubbard}
H_{{\rm film}}=\sum_{\ve{k}\sigma}\varepsilon_{\ve{k}}
c_{\ve{k}\sigma}^{\dag}c_{\ve{k}\sigma} +U\sum_i n_{i\uparrow}
n_{i\downarrow}+V \sum_{\langle ij \rangle} n_i n_j
\end{eqnarray}
Here $c_{\ve{k}\sigma}^{\dag}$ are electron creation operators,
$n_i=\sum_{\sigma} n_{i\sigma}$ and $V$ is the nearest-neighbor Coulomb
repulsion. Furthermore, to study the charge ordering in the film we introduce
the index $\alpha$ for two sublattices $I$ and $II$ and the order parameter
$\delta=n_I-n_{II}=x_{II}-x_I$ where $n_{I/II}$ are the sublattice electron
concentrations and $n=(n_I+n_{II})/2$. To specify the interface microstructure,
we consider a high-$T_c$ cuprate film where the superconductivity can be tuned
by electrostatic interface doping \cite{mannhart}. For such films with
CuO$_2$--planes between electrically neutral BaO-layers, recent TEM studies
show the most probable interface arrangement to be of a stacking sequence
TiO$_2$--BaO--CuO$_2$ \cite{bals}. This implies that the apical oxygen of the
BaO--layer is shared between the interface TiO$_6$--octahedra and the
CuO$_2$--plane. In the emerging Ti--O--Cu interface
chemical bonding, the Coulomb interaction of the hybridized $3d$--$2p$
TiO-electrons with the charge carriers of the CuO$_2$--planes remains almost
unscreened (about $1$--$2$~eV) and can strongly affect the planar electron
transfer \cite{pavlenko_kopp}. In the context of the interface physics related
to the Ti--O--Cu bonding, we focus on two aspects.

The first ({\it electronic}) aspect concerns the modulation of the $p$--$d$-hybridization
in TiO through doped holes in the CuO$_2$-planes, represented by a term
\begin{equation}\label{pd_hyb}
H_{{\rm pd-hole}}=V_{pd}\sum_i (1-n_i) (p_i^\dag d_i+d_i^\dag p_i)
\end{equation}
The $3d$ and $2p$ states in each interface TiO group with
$H_{{\rm gap}}=(\Delta_{pd}/2)\sum_i(d_i^{\dag}d_i-p_i^{\dag}p_i)$ are separated
by the gap $\Delta_{pd} \approx 3$~eV.
As the effective charge of Ti (about $2.2$--$2.89$~$|e|$ \cite{cohen,heifets})
arises predominantly from the TiO--hybridization and provides up to $0.76$
hybridized electrons per TiO-bond, we assume the one-electron
constraint $p_i^\dag p_i+d_i^\dag d_i=1$ to hold for the TiO--hybridization.
This constraint allows us to introduce a pseudospin formalism through $d_i^\dag
d_i-p_i^\dag p_i=-2s_i^z$; $d_i^\dag p_i+p_i^\dag d_i=s_i^x$. Consequently,
the coupling (\ref{pd_hyb}) can be rewritten in terms of a pseudospin flipping
$H_{{\rm pd-hole}}=V_{pd}\sum_i (1-n_i) s_i^x$ and the gap-term by $H_{{\rm
gap}}=-\Delta_{pd}\sum_i s_i^z$.

As the second ({\it lattice}) aspect of the
interface coupling, we analyze the interaction of polar TiO-phonon modes with
the doped CuO$_2$--holes, which is accounted for by a Holstein approach
\begin{equation}\label{u_s}
H_{{\rm pol}}=\hbar\omega_{TO}\sum_i b_i^{\dag} b_i - \eta_0 \sum_{i}
  (1-n_i) (b_i^{\dag}+b_i)
\end{equation}
The phonon operators $b_i^{\dag}$($b_i$) refer to a particular TiO-mode of
frequency $\omega_{TO}$ [$\hbar\omega_{TO}/t=0.1$]; $\eta_0=\sqrt{\hbar\omega_{TO}E_p}$ is the
\begin{figure}
\epsfxsize=8.5cm {\epsffile{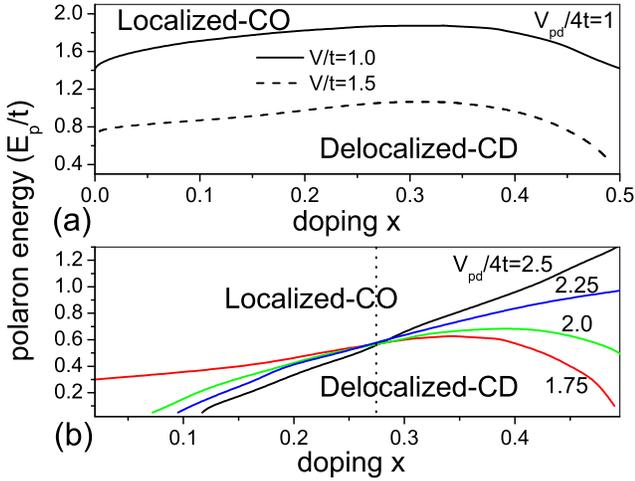}} \caption{Phase diagrams $E_p$--$x$
for localized charge-ordered (CO) and delocalized charge-disordered (CD) states
with different interface coupling $V_{pd}$. Here
$kT/t=0.05$, $U=8t$, $\Delta_{pd}/4t=3$, and $V=1.5t$. In panel (b)
($V_{pd}/4t>1.7$), the dotted line indicates the
crossover between polaron-driven and flipping-driven regimes.
} \label{fig1}
\end{figure}
hole-phonon coupling, and $E_p$ is the polaron binding energy. Applying a
combined two-step unitary transformation $U_{\rm pol}\cdot U_{pd}$ to
(\ref{pd_hyb})--(\ref{u_s}) [$U_{\rm pol}$ is the variational Lang-Firsov
transformation and $U_{pd}=\exp((i V_{pd}/\Delta_{pd})\sum_i s_i^y (1-n_i))$]
\cite{pavlenko_kopp,fehske}, we can eliminate the first-order coupling terms
(\ref{pd_hyb}) in $V_{pd}$ and (\ref{u_s}) in $\eta_0$. Averaging
over the TiO-pseudospin states and phonon bath, we find a
renormalization of the parameters of the extended Hubbard model of the film
including the electron chemical potential $\mu$, Hubbard repulsion and electron
hopping. The Hubbard coupling $U_{\rm eff}=U-\Delta U$ is renormalized by the
attractive contribution $\Delta U=E_{p}\gamma_{\alpha}
(2-\gamma_{\alpha})+V_{pd}^2/4\Delta_{pd}$ \cite{pavlenko_kopp} where the
adiabatic parameters $\gamma_{\alpha}$ for each sublattice should be found by
minimization of the film free energy. The first term in $\Delta U$ contains a
polaronic energy gain whereas the second term comes from the flipping of the
TiO-pseudospins, induced when a neighboring interface hole approaches the
TiO-group. On the other hand, the renormalization leads to a reduction of the
effective inter-sublattice hopping $t_{\rm eff}\sim \xi t$ in the CuO$_2$-plane
by the electron-lattice interface factor $\xi=\xi_{pd}\cdot \xi_{\rm pol}$. The
factor $\xi$ contains the TiO--hybridization contribution $\xi_{pd}=\cos^2
V_{pd}/2\Delta_{pd}$ and the polaron band narrowing factor $\xi_{\rm
pol}=\exp(-\sum _{\alpha}\gamma_{\alpha}^2
E_{p}/2\hbar\omega_{TO}\coth\frac{1}{2}\beta \hbar\omega_{TO})$ where
$\beta=1/k T$. We investigate the thermodynamical stability of the effective
Hubbard model of the film for varying hole doping $x=1-n$ within the
Kotliar-Ruckenstein (KR) approach \cite{kr}. In such an effective model, the
interface parameters $V_{pd}$ and $E_{p}$ modulate the relation between the
electron kinetic and potential energy $V$. Consequently they are responsible
for a transition between the electronic disordered and ordered states of the
film. From the point of view of high-$T_c$ theory, this transition is of
fundamental importance for the film superconducting properties which are
locally suppressed in regions with charge ordering, for example, by a magnetic
field \cite{hoffman}. In the present approach {\it tuning by interfacial
parameters} is considered as a new mechanism to modulate the disorder-order
transition associated with the suppression of superconductivity. In the
superconducting dome the film is non-magnetic. As the interface
coupling with CuO$_2$-holes (electrostatic
coupling with TiO-pseudospins ["dipoles"] and electron-phonon) cannot induce
magnetic moments, we focus on the effect of interface on the
(non-magnetic) electron ordering.

Strong evidence for the feasibility of such an interface-driven tuning is presented
in Fig.~\ref{fig1} with a sequence of phase diagrams $E_{p}$--$x$ for different
values of the coupling $V_{pd}$. In our analysis we choose a weak Coulomb coupling
$V$ ($V/t=1-1.5$) which by itself is not sufficiently strong
to stabilize the ordered state in the film as is seen in Fig.~\ref{fig1}(a):
here the film remains disordered in the entire range of $x<0.5$
and one needs a strong coupling to phonons
$E_p/t > 0.8-1.4$ to stabilize the charge ordering.
The role of the interface becomes even more crucial for
large $V_{pd}/4t>1.7$ (Fig.~\ref{fig1}(b)). Here the curves, corresponding to a
transition between the charge ordered (CO) and disordered (CD) states, cross
approximately in a point which is indicative of
a crossover between two qualitatively different regimes.
In the first, {\it polaron-driven regime}, which persists for lower doping $x$,
the physics in the film is dominated by the electronic kinetic energy.
The latter is suppressed by the factor $\xi$ which is reduced when the coupling to
phonons or TiO-electrons increases. It is essential that
such a suppression of the kinetic energy not only leads to
the localization of holes but is also accompanied by their charge ordering.
These results can be
understood from the analysis of the free energy density
\begin{eqnarray} \label{free}
F=-V(4n^2-\delta^2)+\sum_{\alpha=I,II} [U d_\alpha^2-2\lambda_\alpha^{\sigma}(p_\alpha^2+d_\alpha^2)]\\
-(4/N)kT\sum_{\ve{k},\alpha}
\log(1+\exp(-\beta\varepsilon_{\ve{k}}^\alpha))+2\mu n \nonumber+\Delta E_{\rm IFC},
\end{eqnarray}
where the quasiparticle energies $\varepsilon_{\ve{k}}^\alpha$ form two subbands,
split by the inter-sublattice transfer $t_{\rm eff}$
with gap $\Delta \varepsilon_{I,II}=\sqrt{(\Delta \xi_{I,II}^0)^2+(4t_{\rm eff})^2}$.
Here $\Delta \xi_{I,II}^0=-4V\delta
+\lambda_I^{\sigma}-\lambda_{II}^{\sigma}$ and the Lagrange multipliers $\lambda_\alpha^{\sigma}$
constrain the fermion sublattice occupancies in the KR\ approach. The parameters
$d_\alpha^2$ ($p_\alpha^2$) are the auxiliary boson fields on doubly
(singly) occupied sites. At low $kT/\Delta\varepsilon_{I,II}\ll 1$,
a numerical analysis shows that because of the weak
renormalization of $t_{\rm eff}$ by $\xi \approx 1$ in the disordered state
and because $(\Delta \xi_{I,II}^0/t_{\rm eff})^2\gg 1$ holds in the ordered state, the direct
influence of the interface coupling on the gap $\Delta \varepsilon_{I,II}$ and on all related
terms in $F$ is insignificant. Instead, we find that the low-$T$ interface physics is
contained in $\Delta E_{\rm IFC}$.
The term $\Delta E_{\rm IFC}= -\sum_{\alpha} (\Delta E_{\rm pol}+\Delta E_{pd})$ originates from the
effective interaction $\Delta U$ and adiabatic interface TiO--distortions \cite{fehske} and
consists of the two (polaron and pseudospin) parts
\begin{eqnarray} \label{f_attr}
&& \Delta E_{\rm pol}=-E_p\left \{\gamma_\alpha (2-\gamma_\alpha)y_\alpha^2+
      (1-\gamma_\alpha )^2x_\alpha^2 \right \}\\
&& \Delta E_{pd}=-\left \{ {V_{pd}^2}/{4\Delta_{pd}}\right \} y_\alpha^2
\nonumber
\end{eqnarray}
Each part in (\ref{f_attr}) depends on the spatial hole inhomogeneity, expressed by
$\delta$, through the coefficients $y_\alpha^2=e_\alpha^2+d_\alpha^2$ ($e_\alpha^2=d_\alpha^2+x_\alpha$)
and the polaron adiabatic factors $\gamma_\alpha$;
both are different in the ordered and disordered states.
In the CO-state with one of the sublattices preferably occupied by charge, so that $n_I \sim 2n \gg
n_{II}$, the number of doubly occupied
and empty sites at low $x$ is larger than in the disordered state which
implies $(y_\alpha^{CO})^2>(y_\alpha^{CD})^2$.
The polaron part $\Delta E_{\rm pol}$ in (\ref{f_attr}) has two
competing contributions. The first is proportional to $y_\alpha^2$ and is significant
for $\gamma_\alpha \rightarrow 1$ (small polaron regime).
Conversely, the second term in $\Delta E_{\rm pol}$ is proportional to $x_\alpha^2$ and contributes mostly for weak
polaron coupling $\gamma_\alpha \ll 1$. If the charge order (CO-state) stabilizes for low
$x$, the first, ``small-polaron'' localizing term prevails which immediately leads
to a strong polaron localization with
$\gamma_\alpha^{CO}>\gamma_\alpha^{CD}$. Moreover, in the
CO-state with $n_I\gg n_{II}$ we find a stronger polaron effect for sublattice
$I$: $\gamma_I^{CO} \gg \gamma_{II}^{CO}$.

In distinction to the CO-state, in the disordered state with the dominant
second term in $\Delta E_{\rm pol}$, $\gamma_I^{CD}=\gamma_{II}^{CD} \ll 1$,
the polaron effect is weak and the holes are delocalized on the lattice. As the
strong polaron effect is associated with charge ordering, the suppression of
the kinetic energy by increasing $E_p$ or $V_{pd}$ in the polaron-driven regime
leads to an extension of the stability of the CO-state (Fig.~\ref{fig1},
$x<0.27$) \cite{warning}. We note that for the range of the parameters considered in this work,
the CO-state is found to be stable against the charge separation.

\begin{figure}[t]
\epsfxsize=8.5cm {\epsffile{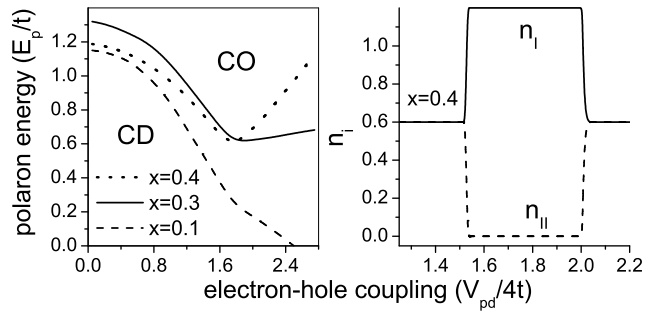}} \caption{Phase diagrams $V_{pd}$--$E_p$ for $kT/t=0.05$
in polaron-driven ($x=0.1$) and flipping-driven ($x=0.3;0.4$) regimes. Right panel: the reentrant CD/CO/CD
transition for $x=0.4$.}
\label{fig2}
\end{figure}

By contrast, in the {\it flipping-driven regime}, which prevails for higher
doping, $x>0.27$, the dominant contribution to $\Delta E_{\rm IFC}$
(\ref{f_attr}) comes from the term $\Delta E_{pd}\sim -(V_{pd}\,y_\alpha)^2$
originating from the effective hole attraction due to the interface
hole-pseudosping coupling. Here the numerical studies show that
$(y_\alpha^{CO})^2<(y_\alpha^{CD})^2$ and consequently the effective potential
energy gain $\Delta E_{pd}$ is larger in the disordered state where the
pseudospin flipping is caused by weakly dressed ($\gamma_\alpha \ll 1$) holes.
This flipping-induced gain leads to a principally new effect when the
disordered (CD) state extends towards larger $E_p$ with an increase of $V_{pd}$
(Fig.~\ref{fig1}(b)).

The direct consequence of the competition between the polaron- and flipping-driven regimes
is clearly seen in Fig.~\ref{fig2}. For lower $x$, the
monotonic decrease
with $V_{pd}$ of the CO/CD-transition curve is caused by the interface-induced hole localization.
Conversely, for larger $x$, the nontrivial behavior with a minimum at $V_{pd}/4t\approx 1.75$ originates from
the competition between the two different types of the interface coupling
[phonon-induced hole localization versus flipping-driven delocalization]. For the considered range of coupling
energies, we obtain significant corrections $-\Delta U$ and  a strong reduction of the Hubbard
repulsion to $U_{eff}\approx 4t$ at the interface. Such a weak $U_{eff}$ is well below the Mott transition and is
not sufficient to prevent high sublattice densities $n_{I}>1$ in the CO state in Fig.~\ref{fig2}.

For weaker interface coupling, $V_{pd}/4t<1.5$, as we approach the range $x\approx 0.5$, the
region of the CO-state extends again (Fig.~\ref{fig1}) which leads to the checkerboard ordering
at substantially smaller $E_p/t \ll 1$. Although the order in the vicinity of quarter-filling
is widely discussed in the literature where it appears due to the nearest-neighbor Coulomb
interactions \cite{bulla2}, we should stress that in our studies, {\it it is the interface coupling,
which plays a key role in the charge ordering of correlated films}.

\begin{figure}[t]
\epsfxsize=8.cm {\epsffile{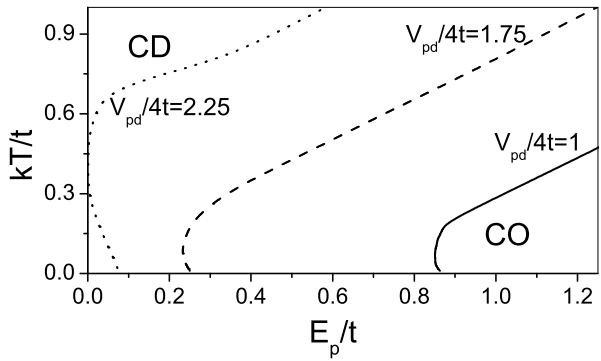}} \caption{Phase diagrams $T$--$E_p$
in the polaron-driven regime at lower doping $x=0.1$ for
different $V_{pd}$.} \label{fig3}
\end{figure}

The main interface factor, which drives the transition for weak Coulomb interaction $V$,
is the coupling to phonons which is clear from Fig.~\ref{fig1} where the
CO-state stabilizes with increasing $E_p$. Moreover, the phase diagrams
$T$--$E_p$ in the polaron-driven
regime (Fig.~\ref{fig3}) show some analogy with the $T$--$V$ diagrams
\cite{bulla2}. The reentrant temperature behavior for $kT/t<0.1$ with the
negative slope of $T_c(E_p)$ can be understood from the analysis of the entropy
contribution $S \sim \sum_{\ve{k},\alpha} \log (1+\exp(-\beta
\varepsilon_{\ve{k}}^\alpha))$ to the free energy $F$. In the CO-state, where the
charge is localized and $t_{\rm eff}/t\ll 1$, the two energy subbands
$\varepsilon_{\ve{k}}^\alpha$ degenerate into two energy levels
$\varepsilon^\alpha=-\mu+4Vn_\alpha+\lambda_\alpha^{\sigma}$. At low $T$, the
entropy $S$ for such $(N/2)$--degenerated two-level systems grows stronger
than the entropy in the disordered state with $\varepsilon_{\ve{k}}^\alpha$ split by
$t_{\rm eff}\sim t$. On the other hand, with a further increase of $T$, the increasing
charge entropy of the homogeneous state dominates
and we have a standard melting behavior.
As $V_{pd}$ increases, the entire curve $T_c(E_p)$ in Fig.~\ref{fig3} is shifted
to smaller $E_p$ which is explained by the $V_{pd}$-induced suppression of
the disorder in the polaron-driven regime.

\begin{figure}[h]
\epsfxsize=8.5cm {\epsffile{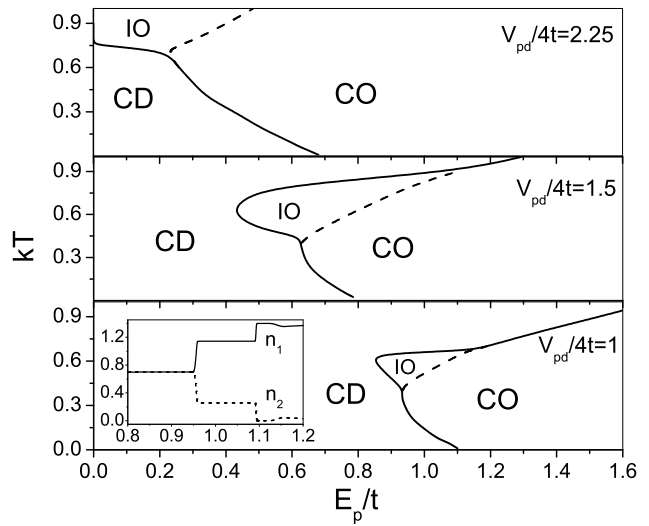}} \caption{Phase diagrams $T$--$E_p$
in the flipping-driven regime for $x=0.3$. The panels show the
appearance of the intermediate IO-phase and the expansion of charge-ordered
(CO) and IO-states upon an increase of $V_{pd}$. Inset:
evolution of $n_\alpha$ through the phase transitions CD-IO-CO with
$E_p$.} \label{fig4}
\end{figure}

There is, however, a crucial difference between the
reentrant behavior in the polaron-driven
and the flipping-driven regime, demonstrated for $x=0.3$ in Fig.~\ref{fig4}.
Here, in the reentrant area, we find a new, intermediate-ordered (IO) state
[indicated by dashed curves in the $T$--$E_p$ phase diagrams]. In distinction
to the disordered phase, the I/II-concentration difference
$\delta$ in the IO-state is less than in the CO-state (Fig.~\ref{fig4}, inset).
Moreover, contrary to the CO-state where the charge in the sublattice $I$ is
strongly localized, in the IO-state we find a rather weak polaron effect with
$\gamma_\alpha^{CD}<\gamma_\alpha^{IO}<\gamma_\alpha^{CO}$. Due to $\delta^{IO}<\delta^{CO}$,
the stability of this intermediate state can be explained by the smaller gap $\Delta\varepsilon_{I,II}$
and consequently larger $S$-contribution in expression (\ref{free}) for $F$.

In conclusion, we have studied the
interface tuning of electronic states
in a correlated cuprate film through degrees of freedom which orginate from an interface dielectric layer.
Even for a weak nearest-neighbor Coulomb coupling, the
doped charge in the film
can be ordered and localized by the interface coupling,
which results in the suppression of superconductivity.
We identified a competition between two
types of coupling (to phonons and to hybridized electrons). It becomes manifest
in a crossover between the underdoped regime with a dominating charge ordered (localized) state,
and the high doping regime where a state of almost free mobile holes stabilizes.
With the emerging
possibilities for electrostatic and chemical control of interface couplings, the considered interface tuning
of the order-disorder transition can be essential
for controlling superconductivity in the film.

This work was supported through the DFG~SFB-484, BMBF~13N6918A,
and DAAD D/03/36760.


\begin{references}
\bibitem{warning} The mean-field analysis of a simplified extended Hubbard model with
a similar interface hole-phonon coupling term leads to a suppression of CD-state and to a polaron-driven
charge localization, in agreement with the present studies. However, more careful studies would require
an extensive analysis of the charge fluctuations and disorder which can substantially affect the
interface coupling to phonons and excitons.
\end{references}
\end{document}